\documentstyle[12pt]{article}

\newcommand{\EQ}{\begin{equation}}
\newcommand{\EN}{\end{equation}}
\begin{document}
\topmargin 0pt
\oddsidemargin=-0.4truecm
\evensidemargin=-0.4truecm

\renewcommand{\thefootnote}{\fnsymbol{footnote}}

\newpage
\setcounter{page}{1}
\begin{titlepage}
\begin{flushright}
ICTP Ref. IC/93/31\\
February 1993
\end{flushright}
\vspace*{0.4cm}
\begin{center}
{\LARGE\bf CHARGED SCALAR PARTICLES AND\\
$\tau$ LEPTONIC DECAY }
\vspace{0.8cm}

{\large\bf
Zhijian Tao}

\vspace*{0.8cm}

{\em  International Centre for Theoretical Physics\\
\vspace*{-0.1cm}
Strada Costiera 11, I-34100 Trieste, Italy\\}
\end{center}
\vspace{.2truecm}

\begin{abstract}
Charged scalar particles introduced in some extensions of
the standard model can induce $\tau$ leptonic decay at tree
level. We find that with some charged  SU(2)-singlet scalar
particles, like ones introduced in Zee-type models, $\tau$
leptonic decay width is always smaller than what  is predicted
by the standard model, therefore they may offer a natural solution
to $\tau$ decay puzzle. To be more specific, we examine some
Zee-type models in detail to see if at the same time they are
acceptable in particle physics, cosmology and astrophysics.
It is shown that $\tau$ decay data do put some constrains on
these models.
\end{abstract}
\vspace{2cm}
\centerline{PACS numbers: 13.35.+s, 13.10.+q, 12.15.Ff}
\vspace{.3cm}
\end{titlepage}
\renewcommand{\thefootnote}{\arabic{footnote}}
\setcounter{footnote}{0}
\newpage
$\tau$ lepton is an interesting system to test
the standard model (SM) and search for new physics,
since $\tau$ is the heaviest lepton yet known.
There is a long-standing puzzle in $\tau$ lepton decays,
which has received attention for some
years. The puzzle is that the measured $\tau$ lifetime may be
longer than one expected in the SM with three families
\cite{Perl,Marc}. From Particle Data Group (PDG) \cite{Particle},
the measured $\tau$ lifetime is
$\tau^{exp}=(3.05\pm 0.06)\times 10^{-13}s$,
while the SM's expectation  is
$\tau^{th}=(2.87\pm 0.07)\times 10^{-13}s$, where
$m_{\tau}=1784.1\begin{array}{c}+2.7\\ -3.6\end{array}$ MeV is used.
So one sees that the measured $\tau$ lifetime is about $2.3\sigma$
higher than SM expectation value \cite{Marc}. The latest
measurement of $\tau$ mass at BES \cite{BEPC} $m_{\tau}=1776.9\pm
0.5$ MeV
somehow relaxes the $\tau$ lifetime problem. But this downward
shift of $m_{\tau}$ is not enough. $\tau^{exp}$ is still about
$1.9\sigma$ higher than $\tau^{th}(=2.92\pm 0.04\times 10^{-13})$ .
Of course it is very possible that this $\tau$ decay puzzle will
disappear when new  measurements
of leptonic decays become available,
as the expected value does not deviate too much from the measured
one (In fact, it is noticed that there are some new measurements
after PDG \cite{Particle} on $\tau$ lifetime and leptonic decay
branching ratios.
We will comment on that at the end of the paper). However, we feel
that there are some theoretical motivations for taking this puzzle
seriously, such as the existence of a fourth generation or
charged scalar particles (the latter case will  be discussed in
detail later).

One simple solution to $\tau$ decay puzzle is to introduce a fourth
generation \cite{Marc,Li}. If the mixing between $\tau$ neutrino
$\nu_{\tau}$ and the fourth heavy neutrino (it must be heavier than
45.3 GeV from LEP Z-width constraints \cite{LEP}) is around
$sin^2\theta_{mix}\simeq 0.05$, the central value of $\tau^{exp}$
is in consistency with the corresponding theoretical expectation
value in this model.  We denote $g_e$, $g_{\mu}$ and $g_{\tau}$ as weak
couplings of $e$, $\mu$ and $\tau$ leptons respectively,
universality of weak interaction  means $g_e=g_{\mu}=g_{\tau}$.
Any deviation from this relation (for example $g_e=g_{\mu}\not=
g_{\tau}$) implies a violation of the universality (of $\tau$).
In the four generation model, $\tau$ universality in the charged
and neutral current sector is obviously violated by a small amount
\cite{Ma}.
Nevertheless, this is not favored
by the neutral-current data from Z decay which agrees with the
universality of the weak interaction of $e$,
$\mu$ and $\tau$ leptons at the level of precision better than
$0.5\%$ \cite{Ohio}.
Another simple solution assuming a mixing of $\nu_{\tau}$ with
a singlet  neutrino $\nu$ \cite{Aguila} is also in conflict with
this Z decay data.
In addition, following Marciano's analysis \cite{Marc} but using
$m_{\tau}=1776.9\pm 0.4\pm 0.3$ MeV,  we  see that for $\tau$
semileptonic decays $\tau\rightarrow\nu_{\tau}\pi/K$ and
$\tau\rightarrow\nu_{\tau}\pi^-\pi^0$ experimental values agree
with the SM's prediction very well (within $1\sigma$).
So this is another evidence supporting  $\tau$ universality in
semileptonic decay.

In this letter we discuss the effects induced by  scalar particles
in $\tau$ leptonic decay.  Without losing generality we will consider
SU(2)-singlet, doublet and triplet scalars. Generally, for a
SU(2)-singlet or triplet scalar $h_{ab}$, the couplings between
scalars and leptons may be introduced as
$\Delta L_y=f_{ab}l_al_bh_{ab}$, for SU(2)-doublet scalars
$\Delta L_y=f_{ab}{\bar l_a}e_{b}^Ch_{ab}$, where $l_a$ is lepton
doublet, $e_{b}^C$ is lepton singlet and a, b denote family indices.
Due to the fermi statistics, $f_{ab}$ is antisymmetric and symmetric
for singlet and triplet scalars respectively. While for doublet scalar
there is no any constraint on the structure of $f_{ab}$. With these
interactions, it is easy to see that for $\tau$ leptonic decay, for
 example $\tau\rightarrow\mu\nu_{\tau}{\bar\nu_{\mu}}$, two processes
 contribute at tree level. One is SM's W-boson exchanging process, the
other is through exchange of  the $h_{ab}$ particle  (Fig. 1).
 The general features we find are that for singlet $h_{ab}$, the
interference term between these two processes are always negative;
for triplet $h_{ab}$, if $a\not=b$, the interference term is always
positive, if $a=b$ and $f_{aa}f_{bb}>0$ the interference term is
negative. These are because of the antisymmetric and symmetric
 properties of $f_{ab}$. As for the doublet, if we take all of
the fermion masses in final states as zero, then the interference
 effect vanishes. This is  a result of the fact that the Yukawa
coupling   changes the chirality of the leptons, and therefore the
 interference term must be proportional to the charged lepton mass
in the final states. Consequently, we see that with the singlet or
 triplet scalar couplings (with $a=b$ and $f_{aa}f_{bb}>0$) $\tau$
 leptonic widths are smaller than that predicted in the SM, with
 triplet scalar couplings (in the case of $a\not=b$) or doublet
 scalar couplings $\tau$ leptonic decay widths are larger than
that in the SM. Hence $\tau$ decay data suggests that one should
consider  models with singlet or triplet
(with $a=b$ and $f_{aa}f_{bb}>0$) scalars. We show that the
 scalar particle introduced in Zee-model has exactly the required
 property . So as a good example, we will discuss $\tau$ decay puzzle
concentrating only on Zee-type models.

  Zee model was proposed to generate Majorana neutrino masses
 \cite{Zee}.
Recently, it was  found that it can generate large neutrino
 transitional magnetic moment, while keeping neutrino masses
small naturally \cite{BFZ}. Zee-type models are also proposed
 to incorporate  neutrinos with mass at the  order of KeV \cite{BH}
 and at the same time to give a solution to solar neutrino problem (SNP)
\cite{Akh}. In this sort of models, some charged scalar particles $h$ are
introduced,
which carries two units of lepton charge.  The point is that the
$\tau$ leptonic decay widths in these models are  always smaller
than that predicted in SM. Also $\tau$ universality in the
neutral current sector is not violated. On the other hand,
since $h$ does not couple to quarks, we  expect the $\tau$ universality
 is well respected in semileptonic decays.
These are in perfect agreement with the Z decay and $\tau$ leptonic
decay data as well as the more precisely measured value
$|g_{\mu}/g_e|=1.0031\pm 0.0023$ obtained from $\pi$ decay
 \cite{Britton}.

 $h_{ab}  (a\not=b)$  couple to leptons as
\EQ
\Delta L_y=1/2f_{ab}l_a^TCi\tau_2 l_bh_{ab}+h.c.
\EN
where  C is the Dirac charge conjugation matrix and  $f_{ab}$ is
antisymmetric due to fermi statistics. $h_{ab}$ carries two units
 of lepton numbers ($L_a, L_b$). One sees that this interaction has
a global $U(1)_e\times U(1)_{\mu}\times U(1)_{\tau}$ symmetry. So
lepton numbers $L_a$ are not violated through this interaction. There
 are three independent couplings $f_{e\mu}$, $f_{e\tau}$ and
$f_{\mu\tau}$. We assume that $f_{e\mu}$ is considerably smaller
than $f_{e\tau}$ and $f_{\mu\tau}$, so that we don't need to readjust
 fermi constant $G_{\mu}$ ( in fact one or two order of magnitude
smaller is enough, since we don't want to fine-tune
 $f_{e\mu}$ either). This is also  consistent with the constraint
set by universality between beta and $\mu$-decay \cite{Zee,Britton}.
The leptonic decay width (including electroweak radiative corrections
 of the SM) in Zee-type models reads
$$
\Gamma_l=\displaystyle{\frac{G^2_{\mu}m_{\tau}^5}
{192\pi^3}[f(\frac{m_l^2}{m_{\tau}^2})
(1+\frac{3m_{\tau}^2}{5m_W^2})(1+\frac{\alpha(m_{\tau})}
{2\pi}(\frac{25}{4}-\pi^2))-C_{h}]}
$$
\EQ
C_{h}=3.0f_{l\tau}^2(100{\rm GeV}/m_h)^2
\EN
here $f(x)=1-8x+8x^3-x^4-12x^2\ln x$, $G_{\mu}$ is fermi coupling
 constant and $\alpha$ is fine structure constant.
If we consider the weak scale as the only physical scale in
 this work,
i.e. $m_h\simeq 100$ GeV, and take $f_{l\tau}=0.12$, $\Gamma_l$ will
 be about 4\% smaller than what predicted in SM. Consequently $\tau$
 lifetime is about 4\% longer than SM's prediction as the central
 value of $\tau^{exp}$ implies.  It is interesting that with a
reasonable choice of coupling constant $f_{l\tau}\sim 0.1$, $\tau$
lifetime is about a few percent longer than the SM's prediction.
This is a general prediction of Zee-type models. Since $x$ is
 proportional to $f_{l\tau}^2$, a reduction of  $f_{l\tau}$ by
one order of magnitude will decrease $x$ by two order of magnitude.
 In other words, $f_{l\tau}$ can be determined very well from
$\tau$ lifetime measurement, if  $\tau$ lifetime is indeed
 different from the SM's prediction. A few percent deviation
between experimental measurement and the SM's expectation
implies $f_{l\tau}\sim 0.1$ and an inconsistency at $10^{-3}$
level corresponds to $f_{l\tau}\sim 0.02$. In addition, $f_{ab}$
is directly related to neutrino masses, mixing and magnetic
 moments. Precision measurement of $\tau$ lifetime will
constrain these
parameters and then affect the descriptions
of some phenomena of cosmology and astrophysics. In order to
see how well  the models  works when the parameters are fixed from
the $\tau$ decay, we examine some of the aspects which are sensitive
 to the values of $f_{ab}$  in some concrete examples.

{\em BFZ model} \cite{BFZ}\hskip 1cm In this model both neutrino
masses and magnetic moments are generated at two loop level,
but the neutrino masses are suppressed by a factor
proportional to the mass square of the charged leptons due to
the spin suppression mechanism. Individual lepton number is
certainly violated because of the non-zero neutrino transitional
 magnetic moments and masses.
This violation  is only due to  the scalar interaction \cite{Tao}
\EQ M_{ab}^{\alpha\beta}h_{ab}^+(\phi^-_{\alpha}\phi_{\beta}^0-
\phi^-_{\beta}
\phi_{\alpha}^0)
\EN
here $\phi_{\alpha}^-$, $\phi_{\alpha}^0$ belong to Higgs doublets.
The neutrino mass matrix for three lepton flavors reads
\begin{equation}\left(\begin{array}{lll}
0 & m_{e\mu} & m_{e\tau} \\{}
m_{e\mu} & 0 & m_{\mu\tau} \\{}
m_{e\tau} & m_{\mu\tau} & 0
\end{array}\right)
\end{equation}
and  $m_{ab}\propto f_{ab}(m_a^2-m_b^2)$, so one has
$m_{e\mu}\equiv m<<M\equiv m_{e\tau}\sim m_{\mu\tau}$. Therefore
this mass matrix has an approximate symmetry $L_e+L_{\mu}-L_{\tau}$.
 The eigenvalues of this mass matrix are $m_1\sim m,
\hskip 0.4cm m_{2,3}\sim \sqrt{2}M\pm m/2$
indicating the mixing angle between $\nu_e(\nu_{\mu})$ and
 $\nu_{\tau}$ to be    order of $m/M$, and $\nu_e$, $\nu_{\mu}$
 are mixed with large mixing angle close to $45^0$.
Taking $f_{e\tau}\sim f_{\mu\tau}\sim 0.1$ and $f_{e\mu}\leq 10^{-2}$
to satisfy $f_{e\mu}<<f_{e\tau}, f_{\mu\tau}$, one has neutrino
 transitional magnetic moments $(d_{\nu})_{e\tau}\sim
 (d_{\nu})_{\mu\tau}\sim 10^{-11}\mu_B$ and  $(d_{\nu})_{e\mu}\leq
10^{-12}\mu_B$.
With $m$  smaller than $10^{-4}$ eV, and $M$ as large as 0.1  eV,
it is easy to see that the relevant
squared mass difference  $(\Delta m^2)_{e\mu}\simeq M^2\simeq
10^{-2}$ eV$^2$ in $\nu_e$ and $\nu_{\mu}$ oscillation is too large
 and $(d_{\nu})_{e\mu}\leq 10^{-12}\mu_B$ is too small
so that it can not provide a solution to SNP either through
neutrino
oscillation or spin-flavor precession between $\nu_e$ and
 $\nu_{\mu}$.
However, we notice that the oscillation between $\nu_e$ and
 $\nu_{\mu}$ can be responsible for the recently reported
 deficiency of atmospheric $\nu_{\mu}$ (ANP) \cite{Kami}.
Because the squared mass difference
and mixing angle perfectly fit the required parameter
range \cite{Kami}.
As for the SNP we have a solution resorting to  non-resonant
 spin-flavor precession  between $\nu_e$ and $\nu_{\tau}$
when $\nu_e$ goes through  magnetic field inside the Sun.
The reason is that the magnetic moment $(d_{\nu})_{e\tau}$ can be
 sufficiently  large $\sim 10^{-11}\nu_B$ and
 $(\Delta m^2)_{e\tau}\simeq Mm\leq 10^{-5}$ eV$^2$ is
as small as required \cite{Pal}.  Of course
we should also check on  all of the experimental constraints
from particle physics, cosmology and astrophysics. As there are
 only three light
 neutrinos in our case, it seems that there is no problem with
the limits from cosmology and astrophysics. Neutrino oscillation
experiment gives some restrictions on mixing angle and mass
 difference between two neutrino species.  Because of the
approximate symmetry $L_e+L_{\mu}-L_{\tau}$ the only possible
disagreement with experimental constraints could happen in
$\nu_e$ and $\nu_{\mu}$ oscillation. Given a large mixing angle,
neutrino oscillation experiment requires
$(\Delta m^2)_{e\mu}<0.09$ eV$^2$ \cite{Particle},
so our results agree with this restriction. However,
if we take a much more stringent limit
 $(\Delta m^2)_{e\mu}<1.5\times 10^{-3}$ eV$^2$ \cite{Berger},
then our prediction is in conflict with this limit. Surely we
have some freedom to tune the parameters to make $M$ a few times
 smaller
in order to satisfy  this limit.
But at the same time we also reduce magnetic
moment $(d_{\nu})_{e\mu}$ by a same factor. This is not what
we would like to do.
Another stringent constraint comes from the measurement
of rare $\mu$ decay $\mu\rightarrow e\gamma$ \cite{Particle}.
In present case this decay  can happen
but is dominated by two loop diagrams \cite{Dia2},
as lepton number is violated only through scalar interaction (3).
A rather conservative estimate gives $f_{e\mu}^2< 10^{-2}$, this
 is consistent with our requirement $f_{e\mu}<<10^{-1}$.

{\em BH model} \cite{BH} \hskip 1cm This model is based on a global
 lepton flavor symmetry $G=U(1)_e\times U(1)_{\mu}\times
 U(1)_{\tau}$ and G is spontaneously  broken down to
$U(1)_{e-\mu+\tau}$ at weak scale.
Because of the $U(1)_{e-\mu+\tau}$ symmetry,
neutrino mass matrix has the form
\begin{equation}\left( \begin{array}{lll}
0 & m_{e\mu} & 0 \\{}
m_{e\mu} & 0 & m_{\tau\mu} \\{}
0 & m_{\tau\mu} & 0
\end{array} \right)
\end{equation}
$m_{e\mu}$ and $m_{\tau\mu}$ arise from one loop diagrams
\cite{Akh},
giving $M\equiv m_{e\mu}\simeq 1/16\pi^2f_{\mu\tau}
\lambda_{\tau}m_{\tau}$, \hskip 0.4cm
 $m\equiv m_{e\tau}\simeq 1/16\pi^2f_{e\mu}\lambda_e m_{\mu}$,
where $\lambda$'s are scalar coupling constants and we expect
 these to be of the same order of magnitude.  Solving the
eigenvalues of this matrix,
 one gets a massless neutrino which is
 mostly $\nu_e$  mixed with $\nu_{\tau}$ by a mixing angle
$\theta_S\simeq m/M$, and a
 massive ZKM neutrino with mass $\sim M$.
The latest measurement on searching for 17 KeV
neutrino sets a restriction $\theta_S<10^{-3}$ with 95\% CL
\cite{Morrison}. This limit requires $f_{e\mu}/f_{\mu\tau}<10^{-2}$,
 which is again consistent with our assumption on $f_{e\mu}$.
The mass of the heavy ZKM neutrino is naturally about $10\sim
100$ KeV. To avoid any trouble in cosmology and astrophysics ,
this heavy neutrino must decay sufficiently fast and the dominant
 decay is through $\nu_{\tau}\rightarrow \nu_e F$, where $F$ denotes
 flavons associated with the spontaneous breaking of G. The interesting
prediction of this model is also on $\tau$ decay, i.e.
 $B(\tau\rightarrow eF)\simeq 10^{-4}$. With our choice
 of parameters, we get  same value for this  branching ratio.
If we demand the lifetime of the heavy neutrino is less than
$\sim 10^{5}$ sec to be consistent with the conventional
mechanism for large scale structure formation \cite{BH},
then we may put a lower limit $10^{-4}\sim 10^{-5}$ on
 the value of $f_{e\mu}$.  So in this model we can more
or less fix the parameters $f_{ab}$. As the global symmetry
G is spontaneously broken, there may be some additional tree
level $\tau$ leptonic decay processes, like
$\tau\rightarrow\mu{\bar\nu_{\mu}}\nu_e$, but this kind of
 processes are always proportional to the $f_{e\mu}$. Therefore the
corresponding branching ratios are negligibly small.

{\em  Extension of BH model}\hskip 1cm
In order to incorporate SNP and ANP, BH model was extended to
four neutrino species  including one sterile neutrino $n$
\cite{Smi,Akh}. The symmetry group is extended to $G=U(1)_e\times
U(1)_{\mu}\times U(1)_{\tau}\times U(1)_n$. The four neutrino
species may make up one Dirac neutrino and one ZKM neutrino.
Generally there are two possibilities to incorporate SNP. One
is that $\nu_e$ and $\nu_{\mu}^C$ are combined to form a light
ZKM neutrino, $\nu_{\tau}$ and $n$ (or $n^C$) make up a heavy
Dirac neutrino with mass around 10 KeV. Applying  the BFZ
mechanism  on the light ZKM neutrino part, one can generate a
large transitional magnetic moment between $\nu_e$ and $\nu_{\mu}$,
hence providing a spin-flavor precession solution to SNP \cite{Akh}.
 However, in present case  BFZ mechanism does not work  since
 $f_{e\mu}$ is very small, and therefore the expected neutrino
 transitional magnetic moment is not large enough. Another
possibility is that $\nu_{\mu}$ and $\nu_{\tau}$ are combined
to form a heavy ZKM neutrino, and $\nu_e$ and $n$ (or $n^C$) make
up a light Dirac neutrino. Both SNP and ANP can be solved
naturally in this scheme
resorting to the Planck scale effects \cite{Akh,Goran}. As
in BH model the mixing angle  $\theta_S$ between $\nu_e$ and
the heavy neutrino could be naturally smaller than $10^{-3}$
due to the smallness of $f_{e\mu}$.  In addition, with
$\theta_S< 10^{-3}$ the decoupling temperature of the sterile
neutrino $n$ in the early universe is higher than the QCD phase
 transition temperature, consequently the effective number
of neutrino species at the time of nucleosynthesis is smaller
 than 3.3 \cite{Akh}.
Also, there are more $\tau$ leptonic decay channels in this case
 than  in BH model, for example $\tau\rightarrow e{\bar\nu_{\mu}}n$
 or $\tau\rightarrow e{\bar\nu_{\mu}}(n^C)$ could  happen at tree
level. But the amplitudes of the tree diagrams for these  channels
are proportional to light neutrino mass, so they are too small to
 give rise  to any observable effect.

In summary, it is shown from above discussion that in Zee-type
models the branching ratios of the $\tau$ leptonic decays are
 naturally smaller than the predictions of the SM. Therefore
 $\tau$ lifetime
in these models is longer than that in the SM. This suggests a
solution to $\tau$ decay puzzle. Moreover, the universality
 between
e, $\mu$ and $\tau$ is not violated in the neutral current sector
and is also respected in semi-leptonic decays of $\tau$. These are
 favored by current experimental data.

We discussed three Zee-type models in detail, but only
concentrated on the issues which could be sensitive to the
values of $f_{ab}$ fixed by solving $\tau$ decay puzzle.
As to other issues, there are lengthy discussions in original
 papers on these models.
We have analyzed three specific examples to illustrate the
idea of how to incorporate $\tau$ decay puzzle and take into
account other related issues of particle physics, cosmology and
astrophysics. It is very possible that there are some other kind
of  models which can do the same job \cite{Babu}. We  used
Zee-type models as an example, since we think that the models
 discussed above are among the most popular and simplest
 extensions of the SM .

All of the previous analysis are based on the $\tau$ decay data
which implies existence of the  $\tau$ decay puzzle. However,
whether there is really a disagreement between experiment and
the SM on $\tau$ lifetime is  not very clear yet. In this work
we used the data from PDG \cite{Particle}, however, meanwhile
 some new data have become available. In ref. \cite{Ohio}, it
 is reported that the new world average data of $\tau$ decay
agree with the SM within $1\sigma$. Nevertheless the most precise
  measurement among these new experiments on $\tau$ leptonic decay
 from CLEO indicates the discrepancy is still around $2\sigma$
\cite{CLEO,Mar1}.

Anyway, further efforts on precision measurement of the $\tau$
 decay is certainly  very  much desirable. A confirmation of the
 $\tau$ decay puzzle will imply some new physics beyond the SM,
probably as suggested by the models with some singlet  charged
scalar particles, like Zee-type models, or models with triplet
scalars. A negative  result is of course another evidence in
supporting the SM and will furthermore constrain the parameter
space of these models.

\noindent{\bf Acknowledgement}

The author would like to thank G. Senjanovi\'c and
M. Ahmady for very valuable discussions.

\newpage
\centerline{ \Large Figure Caption}
\vskip 2cm

Fig.1 \hskip 1cm Tree level diagrams which contribute
 to $\tau$ leptonic decay with additional scalar particles.


\begin{thebibliography}{1.}
\bibitem{Tao} Here we introduce three
particles $h_{ab}$,
and $L_e$, $L_{\mu}$ as well as $L_{\tau}$ are
 conserved in interaction (4). In contrast,
only one $h$ particle is introduced
in original BFZ model and only total lepton
 number $L_e+L_{\mu}+L_{\tau}$ is respected
by the couplings of $h$ and leptons. However,
the basic idea and ingredient here are same as
that in original BFZ model.
\bibitem{Dia2} The diagrams are similar to
these diagrams generating transitional neutrino
 magnetic moments in ref. \cite{BFZ}. One only
needs to substitute neutrinos by the corresponding
 charged leptons.
\end{thebibliography}
\end{document}